\documentclass[10pt,twocolumn,letterpaper]{article}

\usepackage{iccv}
\usepackage{times}
\usepackage{epsfig}
\usepackage{graphicx}
\usepackage{amsmath}
\usepackage{amssymb}

\usepackage[pagebackref=true,breaklinks=true,letterpaper=true,colorlinks,bookmarks=false]{hyperref}

\iccvfinalcopy 

\def\iccvPaperID{****} 

\ificcvfinal\fi
\begin{document}

\title{\ A Survey of Serverless Machine Learning Model Inference}

\author{Kamil Kojs\\
IT University of Copenhagen, Denmark\\
{\tt\small kako@itu.dk}
}

\maketitle

\begin{abstract}
Recent developments in Generative AI, Computer Vision, and Natural Language Processing have led to an increased integration of AI models into various products. This widespread adoption of AI requires significant efforts in deploying these models in production environments. When hosting machine learning models for real-time predictions, it is important to meet defined Service Level Objectives (SLOs), ensuring reliability, minimal downtime, and optimizing operational costs of the underlying infrastructure. Large machine learning models often demand GPU resources for efficient inference to meet SLOs. In the context of these trends, there is growing interest in hosting AI models in a serverless architecture while still providing GPU access for inference tasks. This survey aims to summarize and categorize the emerging challenges and optimization opportunities for large-scale deep learning serving systems. By providing a novel taxonomy and summarizing recent trends, we hope that this survey could shed light on new optimization perspectives and motivate novel works in large-scale deep learning serving systems.
\end{abstract}

\section{Introduction}
Model serving systems are designed to handle user inference requests in real time. This interactive nature differentiates them from model training systems, which are primarily focused on maximizing data processing throughput. The design of model serving systems is driven by several key goals. Firstly, high performance is crucial, as the system needs to process requests swiftly, even under variable and high-demand workloads. Secondly, cost-effectiveness is a major consideration, as the system should be able to handle a large volume of requests without incurring excessive costs. Thirdly, ease of management is important, allowing data scientists to deploy machine learning models efficiently without being slowed down by intricate details of resource management.

Currently, cloud service providers offer a range of model serving options, including both managed machine learning services and self-managed server solutions. However, these offerings often do not fully satisfy the aforementioned objectives. Available options include serverless deployments that lack GPU support, where billing is based on the duration of system usage \cite{CloudRun} \cite{Lambda} \cite{AzureContainerApps}. This model, while seemingly efficient for non-machine learning applications, can lead to increased latency in response times, potentially causing failures in meeting established SLOs due to its lack of GPU support. On the other hand, there are deployments that provide GPU access for machine learning models, but these typically involve continuous billing throughout the entire lifecycle of the infrastructure \cite{VertexAI} \cite{SageMaker} \cite{AzureAI}. This means that costs are incurred 24/7 for online serving systems, significantly increasing the operational expenses for businesses that offer AI solutions.

Recent developments and trends show that the model parameter sizes are growing substantially with every new generation of developed models. Notable examples include OpenAI's GPT-3.5 and its successor GPT-4 with parameter sizes equivalent to 175 billion and 1.76 trillion respectively \cite{gpt-parameter-count}. The big parameter size and complexity of these models make them unsuitable for environments that do not have access to GPU resources. Generally, any solution that incorporates large deep learning models and necessitates frequent machine learning inference is contingent on the availability of GPU resources. This requirement underscores the critical role of GPUs in the efficient and effective deployment of advanced deep-learning models.

Another critical aspect of deploying machine learning models for inference tasks is the suboptimal utilization of the underlying GPU infrastructure. While GPUs are typically used efficiently during the training process, they are often underutilized during inference. Commonly, in production environments, each GPU is dedicated to a single model at a time, leading to a situation where a significant portion of the GPU's capacity remains idle. Therefore, businesses are still required to bear the cost of the entire provisioned infrastructure, even though it is not fully utilized. Currently, there are limited tools available that support the simultaneous hosting of multiple models on a single GPU. In practice, businesses often need to deploy multiple machine learning models concurrently to meet their predictive serving needs. Due to the scarcity of tools that facilitate the deployment of multiple models on a single GPU, these businesses are forced to allocate each model to a separate instance. This approach leads to inflated deployment costs and results in underutilization of each instance's GPU capabilities.

Having in mind the current state of model serving systems, it becomes evident that there is a pressing need for more balanced and efficient solutions. The existing services, while offering a range of functionalities, often fall short of addressing the comprehensive requirements of modern AI and machine learning applications. This gap in the market highlights the necessity for innovative approaches that can optimize the use of resources, particularly GPUs while ensuring high performance, cost-effectiveness, and ease of management. In light of these challenges and opportunities, this survey aims to identify the emerging trends and potential solutions in the field of model serving systems. By examining the latest advancements and identifying the key areas for improvement, the survey seeks to contribute valuable insights and propose new directions for the development of more effective model serving frameworks in the rapidly evolving landscape of AI and machine learning.

\section{Methods}
This study incorporates guidelines defined in the "Preferred Reporting Items for Systematic Reviews and Meta-Analyses" (PRISMA) statement. The application of PRISMA principles to this research ensures the following:

\begin{itemize}
  \item Transparency and Clarity - We report our research methods comprehensively, providing readers with a clear understanding of our approach to data collection, selection criteria, and analysis.
  \item Bias Minimization - We minimize bias by systematically selecting and reviewing relevant literature without cherry-picking studies that align with specific conclusions.
  \item Reproducibility - Our research methods are described in sufficient detail to enable replication or verification of our findings by other researchers.
  \item Quality Assurance - We critically assess the quality of included studies, ensuring that our literature overview/case study is based on reliable and credible sources.
  \item Credibility - Our adherence to PRISMA guidelines enhances the credibility of our work by demonstrating a standardized and rigorous methodology.
  \item Efficiency - We follow a structured framework that streamlines the research process, preventing the oversight of important studies.
\end{itemize}

By employing the PRISMA guidelines, we aim to deliver a literature overview that is methodologically robust, transparent, and trustworthy, providing valuable insights into our research topic.

\subsection{Literature search strategy}

\begin{table*}[t]
\begin{tabular}{|l|l|l|}
\hline
\textbf{Screening phase} &
  \textbf{Inclusion} &
  \textbf{Exclusion} \\ \hline
\textbf{\begin{tabular}[c]{@{}l@{}}Title, abstract, key-\\ \\ words\end{tabular}} &
  \begin{tabular}[c]{@{}l@{}}The study mentiones techniques of serving \\ multiple ml models in serverless manner, \\ or techniques that can be used for serverless \\ approach to hosting ml models for online \\ predictions.\end{tabular} &
  \begin{tabular}[c]{@{}l@{}}- Type of study not of interest - i.e. not related \\ to machine learning models\\ - Study focusing on training of ml models and \\ not on inference/serving part\end{tabular} \\ \hline
\textbf{Full-text} &
  \begin{tabular}[c]{@{}l@{}}The study reports a clear description of at \\ least one technique that can be used for \\ serverless approach to hosting ml models \\ for online predictions.\end{tabular} &
  \begin{tabular}[c]{@{}l@{}}- The study was to be excluded during the first \\ screening phase but exclusion criteria fulfillment \\ was not clear prior to full-text screening\\ - Studies not containing at least one technique \\ that can be used for serverless approach to \\ hosting ml models for online predictions.\end{tabular} \\ \hline
\end{tabular}
\caption{Eligibility criteria including inclusion and exclusion requirements}
\label{table:inclusion-exclusion}
\end{table*}

In order to collect scientific literature related to the researched topic, arxiv.org and scholar.google.com websites were used for the collection process. The search strategy consisted of 3 parts:

\begin{enumerate}
    \item Each website was queried with specific keyword sets in order to retrieve relevant information to our research topic.
    \item Each retrieved paper was subjected to inclusion and exclusion criteria that were defined before the search process started.
    \item For the definition of the final set of relevant papers, we also take into account papers referenced by any of the collected works identifiable as “literature review” and more recent papers suggested by domain experts and published after the search process took place.
\end{enumerate}

During the first part of the collection process, the following filters were defined:

\begin{itemize}
    \item "published during last 5 years" - any papers published more than 5 years ago were discarded, as the 'serverless machine learning inference' field is relatively new and is advancing very rapidly. The exclusion of such papers was necessary to gather only recent advancements in the researched field.
    \item "computer science domain" - retrieved documents had to be within the computer science domain in order to increase recall.
    \item "article types limited to conference papers, articles, reviews, and conference reviews" - such filter was applied in order to limit the number of retrieved documents to a manageable size.
\end{itemize}

On top of that, for an efficient search process, query keywords needed to be defined. The process of defining the queries was iterative and the main focus was to effectively search for articles related to the researched topic with an objective of increasing recall and limiting the amount of unrelated documents. In order to find a good balance between limiting the amount of retrieved documents and maximizing the relevance to the researched topic, the following queries were defined:
\begin{itemize}
    \item \texttt{
    serverless AND \break 
    (machine learning OR ml OR gpu) OR \break
    (serving OR inference)
    }
    \item \texttt{
    gpu AND \break
    (ml OR serving) AND \break
    (inference OR model OR multiple models OR sharing)
    }
\end{itemize}

Each of these queries was submitted to selected database sites in order to retrieve relevant documents about serverless machine learning model inference.

\subsection{Eligibility criteria}
The eligibility criteria of the selection process consisted of three phases. Each phase focused on screening different parts of the retrieved documents:

\begin{enumerate}
    \item In the first phase, retrieved papers were quickly screened to identify potential duplicates and older versions of the same paper. If such articles were identified then the latest version was chosen and all other versions were discarded from subsequent analysis.
    \item In the second phase, the title, abstract, and keywords were analyzed. After screening the above sections, the paper was either included or excluded from the following analysis.
    \item In the third phase, full document text was analyzed in order to determine its relevance to our study.
\end{enumerate}

We believe that the above strategy enabled us to efficiently identify articles that are relevant to our study which allows us to perform an in-depth survey about the researched topic.

The inclusion and exclusion criteria for the second and third phases can be seen in table \ref{table:inclusion-exclusion}.

\subsection{Data extraction}
The data extraction process is a crucial point in defining the mapping dimensions of serverless machine learning inference. There are clearly some dimensions around which the research on serverless machine learning inference is defined:

\begin{enumerate}
    \item strategy for efficient loading/unloading of the models to GPU - any strategy or algorithm that might be used for the goal of efficient and fast loading and unloading of the models in order to allow usage of those models in a "function-as-a-service" manner while minimizing latency times for inference purpose.
    \item Strategy for ensuring optimal allocation of resources when deploying multiple models on GPU - any strategy or algorithm that might be used to achieve the goal of serving machine learning models for inference task in an online manner while optimizing used resources.
    \item techniques allowing for serving ml models in FaaS way - any technique used that allows serving machine learning models in a 'function as a service' manner.
\end{enumerate}

Given these three topics, we focus on extracting relevant data from the proposed work in the candidate set.

\subsection{Taxonomy definition}
To synthesize the following research findings, we will develop a taxonomy of methods for serverless machine learning model inference, drawing upon data items extracted from gathered set of papers. This taxonomy will be structured around three key elements: strategies for efficient loading and unloading of models to GPUs, strategies ensuring optimal allocation of resources when deploying multiple models on GPU, and techniques that enable the serving of ML models in a Function-as-a-Service (FaaS) manner.

Ultimately, this taxonomy will serve as a valuable resource for researchers in the field of machine learning and cloud computing. It will provide insights into the current state of serverless ML model inference, identify potential gaps in the existing research, and suggest directions for future advancements. This structured approach will assist in the classification and evaluation of emerging serverless ML model inference methodologies, allowing the development of more efficient, scalable, and cost-effective solutions in this rapidly evolving domain.

\section{Results}
During the data collection process, the specified databases were searched using predefined queries. This initial search resulted in a total of 1,016 works from both databases, which were potentially relevant to our analysis. The subsequent second screening phase which consisted of title, abstract, and keywords filtering, narrowed this selection down to 57 works of direct relevance. Finally, a comprehensive analysis of the full text of these documents further refined the selection, resulting in 34 works that were identified as relevant and included in our final candidate set.

The synthesis of the final candidate set resulted in three primary areas that define the domain of 'serverless machine learning model inference.' The research within this field can be categorized into three main directions, each of which will be elaborated upon in the following sections of this paper.

\subsection{Strategy for efficient loading/unloading of the models to GPU}
Every service, offering machine learning model inference, encounters varying levels of user traffic. It's a typical scenario where each service experiences traffic spikes influenced by user behavior patterns. The frequency of model usage varies significantly among users; some may engage with their models every few minutes, while others might do so every few seconds. Rather than allocating each model to a separate instance – a strategy that leads to considerable underutilization for models that are infrequently used – a more efficient approach would be to deploy multiple models on a single instance, utilizing the same GPU. This method would involve dynamically swapping models in and out of GPU memory based on demand. When a new request is received for a model not currently loaded into the GPU memory, the system could load that specific model, perform the necessary inference, and then unload it after the request has been handled. This approach would free up space for other models that may require immediate request processing, thereby optimizing resource utilization and responding effectively to fluctuating user demands.

This strategy, however, confronts a substantial challenge due to the significant increase in model sizes in recent years. It has become increasingly common for models to occupy over 10GB of GPU memory. When serving models for inference, specific Service Level Objectives (SLOs) are established to ensure that latency times for request processing do not exceed certain thresholds, a crucial factor in maintaining user satisfaction with the services provided. The primary challenge in this context arises when a model, not currently loaded into memory, needs to be utilized for an inference task. Due to the substantial size of these models, the time taken to load them into the GPU becomes a critical bottleneck. Addressing this issue of loading time is currently one of the most pressing challenges in the field, as it directly impacts the ability to meet the defined SLOs while ensuring efficient and responsive service to users.

A primary strategy for addressing this issue involves maintaining model parameters in close proximity to GPU memory to minimize loading times. A notable solution, as proposed in \cite{Computron}, advocates for caching model parameters in RAM. This approach effectively bypasses the time-consuming process of loading model parameters from disk, which can exceed 10 seconds for larger models. Given that contemporary graphics cards offer up to 80GB of memory and host memory (RAM) can significantly exceed this, reaching capacities of 4TB, it is ideally suited for caching thousands of models, thereby facilitating rapid loading times. By caching model parameters in host memory, the transfer to the GPU for inference tasks is significantly reduced, thereby enabling quicker loading and unloading of models from the GPU, particularly when hosting multiple models on the same computing instance.

The study presented in \cite{Computron} introduces a system that employs a model parallel swapping design. This design leverages the combined CPU-GPU link bandwidth within a cluster to accelerate the transfer of model parameters. Such an approach renders the swapping of large models practical and enhances resource utilization. In their work, a request queue operates on a First-In-First-Out (FIFO) basis, while model swapping adheres to a Least Recently Used (LRU) replacement policy, ensuring efficient management and allocation of resources.

\subsubsection{Treating GPU memory as cache}
The strategy of caching models in host memory significantly reduces loading times, often to just a few milliseconds. However, as highlighted in \cite{Clockwork}, the transfer of data from host to GPU memory (approximately 8.3ms for ResNet50) still tends to be more time-consuming than the actual inference process on the GPU (around 2.9 ms). This discrepancy results in periods of GPU idle time during the transfer phase from host memory to GPU memory. To address this inefficiency, the approach presented in \cite{Clockwork} conceptualizes GPU memory as a form of cache. This methodology allows for frequently or recently used models to bypass the time-intensive loading processes, maintaining them directly in GPU memory. Consequently, only those models that are less frequently required for inference are swapped out, optimizing the use of GPU resources and reducing the instances of time-consuming model loads.

A similar approach implemented in \cite{gpu_faas} employs a caching mechanism for machine learning models in GPU memory, enhancing model inference efficiency and optimizing GPU memory management. When a GPU process initiates, it uploads an inference model and reports latency post-inference to the "Datastore". The GPU Manager, in coordination with the Cache Manager, maintains cached models used by active GPU processes.

For function invocation requests, the GPU Manager checks with the Cache Manager if the requested model is already in GPU memory. A cache hit means the model is preloaded and can be used immediately, while a cache miss prompts the Cache Manager to direct the GPU Manager to upload the required model to the GPU. In case of a cache miss, the Cache Manager, based on the GPU's available memory and the model's ID, determines which models to evict, following the GPU's LRU list for efficient memory management.

The "ModelMesh" framework, as detailed in \cite{kserve}, adopts a similar approach to its predecessors in managing machine learning model inference. It preloads model parameters and utilizes GPU memory as a cache, prioritizing the retention of the most frequently used models. Model swapping is managed by the "Model Serve Controller", which employs an LRU caching strategy for efficient model serving.

\subsubsection{Leveraging PCIe and NVLink for model swapping}
Recent developments in networking hardware have significantly influenced model-swapping frameworks, as highlighted in the study presented in \cite{Computron}. This research proposes that in systems where GPUs are connected to the CPU via independent PCIe links, it is possible to load model parameter shards concurrently. This parallel loading leverages the increased aggregate link bandwidth between the CPU and GPUs. Such an approach enables dynamic swapping during the serving of large models across a group of devices, making it as feasible as it is for smaller models.

The study in \cite{FaaSwap} explores the potential of utilizing NVLink hardware in clusters equipped with multiple GPUs to enhance the model-swapping process. As detailed in \cite{nvidiaNVLinkNVSwitch}, NVLink boasts a bandwidth of 900 GB per second, which is over 7x that of PCIe Gen5. The authors propose prioritizing the use of NVLink for transferring models between GPUs, especially when the requested model is on a GPU currently engaged in computations. This approach contrasts with the slower PCIe data transfer method.

The process outlined in \cite{FaaSwap} begins by checking if the target model is already loaded on an available GPU. If so, it executes the request immediately, avoiding any swapping delays. However, if the model is on a busy GPU, the system schedules a GPU-to-GPU swap, preferably using the fastest NVLink connection available. If this is not possible, the framework defaults to a host-to-GPU swap, targeting GPUs connected to idle or lightly loaded neighbors to minimize PCIe contention. In summary, this proposed framework aims to minimize the interference and overhead associated with model swapping for each request, thereby ensuring low inference latency.

\subsubsection{Overlapping model weights loading and GPU kernel executions}
The research presented in \cite{MOE} introduces a dynamic scheduling strategy focused on efficiently offloading sparse parameters, such as model parameters, to maintain performance. This approach aims to optimize performance by overlapping the transfer of parameters from CPU memory with the inference computation in GPU memory. The study leverages the capabilities of different CUDA streams, allowing for the simultaneous loading of the model from the CPU and computation on the GPU. This concurrent operation effectively reduces latency times for model serving.

The study detailed in \cite{demand_layering} introduces a novel "3-stage asynchronous pipeline architecture" which is designed to enhance model parameter loading time. This architecture ingeniously overlaps read, copy, and kernel operations, enabling the parallel execution of inference tasks alongside model loading. The authors of this study made a key observation: during the inference process, model parameters are utilized sequentially, layer by layer, in a forward pass order. This sequential usage means that at any given moment, only a subset of the parameters is actively used. This insight allows for a dynamic approach, where the next set of parameters is loaded while the current layer's kernel operations are being processed on the GPU. This strategy significantly streamlines the inference process, reducing latency and enhancing overall system efficiency.

A similar approach has been incorporated in \cite{FaaSwap} where model parameters are loaded and unloaded in forward-pass layer order to maximize memory allocation. The executor concurrently swaps model parameters and executes CUDA APIs on those loaded, allowing for parallel model parameter loading and kernel executions.

\subsubsection{Sharing model weights/tensors across the same models and replicas of the same function}
The deployment of multiple similar models offers an opportunity to optimize resource utilization by sharing model tensors across instances of the same function. This approach enables frameworks like PyTorch to construct models more efficiently, utilizing zero-copy techniques. The methodology is detailed in \cite{fast-gshare}, which introduces an IPC-based (Inter-Process Communication) model-sharing mechanism. This mechanism comprises two primary components: the Model Store Library and the Model Storage Server.

The Model Store Library is specifically designed to facilitate the sharing of model tensors between instances of identical functions. This design allows an inference framework, such as PyTorch, to construct the model with zero-copy, significantly enhancing efficiency. On the other hand, the Model Storage Server plays a crucial role in allocating memory on the GPU. It exports the corresponding IPC handler to the Model Store Library, enabling direct access to the inference function. This streamlined process not only optimizes memory usage but also reduces the computational overhead associated with model deployment, leading to more efficient and effective use of GPU resources in machine learning inference tasks.

The concept of parameter sharing across multiple model instances, particularly for memory-intensive parameters that remain constant throughout inference executions, is explored in \cite{gslice}. This approach is grounded in the understanding that these parameters do not vary between different runs of the inference process, presenting an opportunity for optimization.

By implementing a method that facilitates the sharing of these parameters once they are transferred to GPU memory, significant improvements in efficiency can be achieved. This strategy effectively reduces the memory footprint of the inference function on the GPU. Consequently, it enables the scaling and multiplexing of a greater number of models, even on GPU devices with limited memory capacity.

The key advantage of this approach is its ability to maximize the utilization of available GPU memory resources. By sharing and reusing parameters across multiple replicas, it becomes possible to host more inference functions concurrently. This not only enhances the overall capacity of the system but also contributes to more cost-effective and efficient use of GPU machine learning serving systems.

The study presented in \cite{trims} involves decoupling the models from their respective functions, overseen by a Model Resource Manager. This manager enables effective model sharing across various Function-as-a-Service (FaaS) functions. By sharing DL models within a centralized model catalog, the system significantly reduces the overhead associated with model loading. This approach not only decreases the end-to-end latency for DL inferences but also substantially reduces the memory footprint. The key advantage of this system is its ability to maintain a single instance of a model in memory, which can then be concurrently accessed and utilized by multiple users.

\subsection{Strategy for ensuring optimal allocation of resources when deploying multiple models on GPU}
Efforts to rapidly swap models in and out of GPU memory present a highly efficient strategy for hosting multiple models on a single instance, significantly reducing the cumulative idle time compared to deploying each model on separate instances. However, the above strategies primarily focus on optimizing the model loading times to the GPU. An aspect of concurrently sharing GPU resources among multiple models in the above works is overlooked. Consequently, the maximum GPU utilization achievable with this approach is limited to that of the largest model being served. While it's feasible to host thousands of models on a single GPU with these methods, they do not address the aspect of GPU sharing during simultaneous inference requests.

This limitation becomes apparent when multiple inference requests are queued. If only one model can perform an inference task at a time, any concurrent requests must wait, leading to increased latency. Moreover, as model inference typically utilizes only a fraction of the available GPU memory and computing power, there is a significant opportunity for improvement. The ideal solution would be a system that enables concurrent model inferences on the same GPU. Such a system would maximize the GPU's resource allocation while adhering to the latency times defined by SLOs, thereby enhancing overall efficiency and performance.

In the development of a serverless machine learning model inference system, it is important to integrate both rapid model swapping and concurrent model execution strategies. This dual approach is essential for achieving efficient resource utilization on GPUs. By combining the capability to quickly load and unload models with the ability to concurrently process multiple inference tasks, the system can maximize GPU efficiency. This holistic implementation ensures that both aspects work together, leading to an optimized balance of resource allocation and performance in serverless machine learning serving environments.

\subsubsection{Partitioning GPU into multiple smaller vGPUs}
One proposed solution to enhance GPU utilization is the adoption of Nvidia's Multi-Instance GPU (MIG) technology, as detailed in \cite{nvidiaMIG}.
According to \cite{nvidiaMIG} MIG can partition the GPU into as many as seven instances, each fully isolated with its own high-bandwidth memory, cache, and compute cores. This partitioning into multiple virtual GPUs enables significantly higher overall utilization, as it facilitates the simultaneous hosting of multiple models on a single GPU instance. This approach has been adapted by works described in \cite{Paris} \cite{inference_with_spatial_partioning} \cite{d-stack} \cite{reconfigurable-gpu} \cite{knix} \cite{gslice}.

The approach of partitioning GPUs into multiple vGPUs for simultaneous inference of multiple models is discussed in several studies, including those cited in \cite{reconfigurable-gpu}, \cite{knix}, and \cite{gslice}. These studies explore the dynamic reconfiguration of GPU resources to adapt to varying user traffic and system demands. They employ various algorithms, such as greedy algorithms, Genetic Algorithms, and Monte Carlo Tree Search, to periodically adjust the partitioning state of the GPU. This reconfiguration aligns the GPU's state with the current traffic patterns and workload requirements, ensuring efficient utilization of GPU resources for concurrent model inference.

In \cite{Paris} they propose a system combined of two key components: a partitioning algorithm (PARIS) and a scheduling algorithm (ELSA), both tailored for reconfigurable GPUs. PARIS (Partitioning Algorithm) is designed for inference servers with reconfigurable GPUs. It systematically identifies the most suitable heterogeneous set of multi-granular GPU partitions, optimizing for the specific needs of the inference server. PARIS takes into account both the model-specific inference properties (such as the trade-off between latency and GPU utility under a given GPU partition size) and the distribution of batch sizes. This allows for the generation of a diverse set of partitioning granularities and the deployment of an appropriate number of instances for each partition.
ELSA (Elastic Scheduling Algorithm) is a high-performance scheduling algorithm that is aware of the heterogeneity of the GPU partitions. It uses an inference latency prediction model to estimate the SLO slack for each query and determines the most suitable among the heterogeneous GPUs to process the query. ELSA's awareness of heterogeneity aids in maximizing server utilization while minimizing SLO violations.

The proposed GPU partitioning approach in \cite{inference_with_spatial_partioning} creates a layer of configurable virtual GPUs called "gpu-lets". A key component of this system is the scheduler, which uses a performance profiler to monitor request rates and collect performance statistics. These insights inform decisions about the size of gpu-lets and model assignments. The scheduler operates periodically, adjusting to user-specified task specifications and model-specific SLO targets. The system comprises two major elements. "Dynamic Partition Reorganizer" is used to monitor incoming rates and schedule tasks for around 20 seconds, a period determined based on the time it takes to reorganize GPU partitions. This process includes spawning new processes with designated MPS resources, loading necessary kernels and models, and a warm-up phase, all executed in the background to minimize the impact on SLOs. "Backend Inference Executor" receives inference requests, model files, and input data, using the PyTorch runtime to initiate inference executions on the gpu-lets. It leverages MPS-supported resource provisioning capabilities. The system periodically runs an algorithm to determine the ideal MPS partitioning and gpu-let sizes for models, adjusting to changing conditions to optimize resource utilization and minimize SLO violations.

The paper described in \cite{d-stack} introduces D-STACK, a dynamic and fair spatio-temporal scheduler designed to enable concurrent execution of multiple DNNs on a GPU. D-STACK employs a model to estimate the optimal level of parallelism (referred to as the "Knee") that each DNN can utilize on the GPU. This helps in allocating the appropriate percentage of GPU resources to different DNNs. D-STACK's scheduler employs a version of the Earliest Deadline First Scheduling (EDF) algorithm. This approach prioritizes models with the tightest deadlines, aiming to fit and run as many models as possible concurrently on the GPU while meeting each model’s specific SLOs.

\subsubsection{Batching}
Another approach introduced in many works \cite{mlproxy} \cite{dataflow} \cite{Throughput_Maximization} \cite{SMDP} \cite{d-stack} \cite{lazy_batching} is batching. In scenarios where users do not require immediate responses and can tolerate latency times of a few seconds without breaching their SLOs, it is advantageous to accumulate a sufficient number of requests before executing the inference task. This batching strategy enhances GPU utilization by allowing the system to process multiple requests simultaneously. Additionally, it provides opportunities for users with more stringent SLO requirements to perform their inference requests in the intervals between batch processing, thereby optimizing overall system efficiency and resource allocation.

The "MLProxy" described in \cite{mlproxy} introduces a batching approach system, composed of two primary components: the Smart Proxy and the Smart Monitor. The Smart Proxy module is engineered to handle incoming HTTP requests efficiently. It employs a dynamic batching algorithm that groups these requests based on specific criteria. The grouped requests are then forwarded to the designated model for inference task. This forwarding occurs either when the accumulated requests reach a predefined maximum batch size or when a set timeout period expires, ensuring a balance between response time and batch processing efficiency.

The "DNNScaler" system introduced in \cite{Throughput_Maximization} offers an approach to enhance throughput in serverless machine learning model inference by employing either Batching or Multi-Tenancy strategies. This system is composed of two main components: the Profiler and the Scaler. The profiler is responsible for determining the most suitable approach for a given DNN, whether it's Batching or Multi-Tenancy. It conducts real-time, lightweight profiling to first assess throughput improvement using an optimal batch size. The Scaler module focuses on identifying the largest batch size or the number of co-located instances (as recommended by the Profiler) that can maintain latency within the specified SLOs. This module is dynamic and can adjust the batch size during the production phase without manual intervention. Adjustments are necessary to accommodate factors like variations in the input dataset, GPU temperature, and GPU usage frequency. The system also allows users to modify the SLO during runtime, with the batch size being automatically adjusted based on real-time metrics to ensure compliance with the new SLO. This flexibility ensures that the system remains efficient and responsive to changing operational conditions.

The work described in \cite{SMDP} models the service as a batch service queue, where the processing time is dependent on the batch size. The core of this strategy is to frame the design of dynamic batching as a continuous-time average-cost problem, formulated as a Semi-Markov Decision Process (SMDP). The objective is to minimize the weighted sum of the average response time and average power consumption. To find the optimal policy, the problem is transformed into an associated discrete-time Markov Decision Process (MDP) with finite state approximation and discretization techniques.

LazyBatching \cite{lazy_batching} is an SLO-aware batching system that uniquely approaches scheduling and batching at the granularity of individual models. A central feature of LazyBatching is its SLO-aware, slack time prediction model. This model is a critical component of the system's scheduler, enabling it to make intelligent decisions about when and which inputs should be batched together. By considering the slack time - the time available before reaching the deadline of SLO - LazyBatching can optimize the batching process, ensuring that inputs are grouped in a way that maximizes efficiency without compromising on the SLO commitments.

\subsubsection{Other promising directions for resource optimization when deploying multiple models on GPU}
In section 3.1.3, the authors of \cite{demand_layering} highlight a crucial aspect of the inference process: the sequential utilization of model parameters, layer by layer, in a forward pass order. This means that at any given moment, only a specific subset of the parameters are in active use. The paper presents an innovative approach where, by swapping model weights in this sequential order, a significant reduction in GPU memory usage can be achieved. Specifically, their method results in a 96.5\% reduction in memory allocation with an average delay overhead of just 14.8\%.

While the original research was focused on embedded systems with limited GPU memory capacities, the principles and techniques discussed are equally applicable to the serverless inference serving domain. By adopting this layer-by-layer swapping approach, each model's memory allocation can be drastically reduced. This reduction in memory usage per model opens up the possibility of hosting a substantially larger number of models on a single GPU, thereby enhancing the efficiency and scalability of serverless machine learning model inference systems.

The space-time scheduling approach \cite{dynamic_space_time_scheduling} is a dynamic method for optimizing the performance of multiple disjoint DNN graphs on GPUs. Unlike traditional single-tenant optimizers like TVM, Tensor Comprehensions, Halide, GLOW, and TensorRT, which focus on optimizing individual graphs through techniques like kernel-fusion and auto-tuning, space-time scheduling aims to efficiently manage multiple concurrent small kernels from different DNN graphs. This method works by merging these small kernels into a smaller number of larger "super-kernels" that effectively utilize the full capacity of the GPU. This approach helps to avoid the scheduling penalties typically associated with space-only multiplexing methods. The space-time scheduler dynamically schedules these kernels as they arrive, with the potential for caching super-kernels for more stable workloads over time.

The key innovation of this approach is its ability to dynamically batch a large number of kernels, particularly those executing similar matrix multiplication routines, and to interleave CUDA streams effectively. This dynamic batching is scalable and can be applied to various real-world neural network tasks. It is designed to complement the existing ecosystem of DNN graph optimizers, offering a scalable alternative to the manual combination of small kernels within a stream. This makes it particularly effective in scenarios where multiple DNN models are being run concurrently, optimizing GPU utilization and enhancing overall performance.

The research presented in \cite{FaaSwap} introduces a strategy where multiple model functions share GPU runtime, significantly reducing the GPU memory overhead associated with loading models. In this approach, CUDA kernels are preloaded onto the GPU. Typically, each loaded model initializes its own GPU runtime, consuming a portion of the GPU's memory. However, by enabling multiple models to share the same runtime space, this method effectively minimizes the memory overhead. This shared runtime approach not only conserves valuable GPU memory but also enhances the overall efficiency of the system, particularly in environments where multiple models are deployed simultaneously.

The primary objective of the performance optimization discussed in \cite{MAD_MAX} is to pinpoint the most effective combinations of kernels and parallelization techniques. These combinations are specifically chosen to reduce exposed communication and maximize the overlap between communication and computation in large-scale machine learning workloads. This approach aims to enhance the efficiency of these workloads by ensuring that communication processes, such as data transfer, occur simultaneously with computation tasks. This concurrent execution helps in reducing idle times and improving the overall throughput of the system, particularly in environments with heavy data exchange and computation demands.

The research described in \cite{cuSync} introduces "cuSync", a novel fine-grained synchronization mechanism for dependent CUDA kernels, designed to enhance the efficiency of parallel computing tasks on GPUs. This mechanism is particularly useful for machine learning models, which often involve highly parallel computations like matrix multiplication, convolutions, and dropout. These computations are commonly executed on GPUs, by dividing the computation into independent processing blocks, known as tiles. Since the number of tiles is usually higher than the execution units of a GPU, tiles are executed on all execution units in waves. However, the tiles executed in the last wave can under-utilize the execution
units because tiles are not always a multiple of execution units. "cuSync" allows thread blocks from all kernels to be executed on Streaming Multiprocessors (SMs) without needing coarse-grained stream synchronizations. This approach helps in better utilization of the GPU's execution units. It adds fine-grained synchronization among the dependent tiles of every producer-consumer kernel pair. This ensures that dependent kernels are executed in the correct sequence, reducing idle times and improving efficiency. The kernels are invoked on different CUDA streams associated with their respective stages, eliminating the need for stream synchronization between kernels. This approach ensures that dependent kernels are executed in the correct order, leading to more efficient parallel computing tasks and better overall performance.

The research proposed in \cite{jit} introduces a novel Just-in-Time (JIT) compiler. It leverages runtime information, such as the number of concurrent execution kernels and the device context (including typical problem sizes handled by the device). This information is used to dynamically allocate resources, ensuring efficient utilization of the GPU. Individual kernels are retuned to improve spatial multiplexing, enhancing the parallel processing capabilities of the GPU. They are coalesced to improve overall device utilization, ensuring that the GPU's computational resources are fully leveraged. Kernels are also reordered to achieve optimal spatio-temporal packing, which helps in managing the execution of multiple tasks more efficiently while meeting latency constraints. Unlike static compilers like TensorRT, this JIT compiler dynamically adjusts to the workloads running on the GPU. This flexibility allows it to respond in real-time to changing demands and workloads, ensuring optimal performance.

Another work described in \cite{NETFUSE} proposes NETFUSE - a technique for merging multiple DNN models that share the same architecture but differ in weights and inputs. It achieves this by replacing standard operations with more generalized versions, allowing specific sets of weights to be associated with particular sets of inputs. NETFUSE transforms traditional matrix multiplications into batch matrix multiplications, handling a batch of inputs and weight tensors simultaneously. It utilizes grouped convolution for merging, where each output channel is derived from a specific group of input channels, differing from the standard convolution operation. It converts standard layer normalization instances into a single group normalization, breaking up channels into disjoint groups for effective merging.

SPLIT \cite{SPLIT} introduces an innovative approach to optimizing the execution of deep learning models by employing a genetic algorithm for model splitting and a preemption method based on a greedy algorithm. In deep learning models, multiple operators are combined, and their interdependencies can be represented as a directed acyclic graph (DAG). The concept of model splitting involves dividing these longer models into smaller segments or blocks at operator boundaries, enhancing manageability and efficiency.

The genetic algorithm at the core of SPLIT is designed to determine the most effective points for splitting the model. This algorithm strategically selects cut points to divide the model into a predetermined number of blocks, aiming for an even distribution. The primary goal is to create blocks that are as uniformly sized as possible, thereby minimizing the standard deviation of block execution times. This uniformity is crucial for reducing jitter and enhancing the predictability and consistency of model performance. Complementing the splitting algorithm, SPLIT incorporates a preemption strategy based on a greedy algorithm. This method is adept at quickly assessing incoming requests to decide whether preempting current processes will minimize latency. This ensures efficient resource utilization and timely response, making the overall system more responsive and effective.

Finally, the research presented in \cite{fast-gshare} introduces a comprehensive approach to GPU sharing, specifically tailored for Function-as-a-Service (FaaS) deep learning inference. This approach encompasses several key components. FaST-Manager is a GPU sharing mechanism designed to manage GPU resources effectively. It is capable of limiting and isolating GPU resources with a high degree of precision, both in terms of temporal and spatial granularity. This means that the allocation and usage of GPU resources can be finely controlled and adjusted according to specific needs and demands. FaST-Profiler and FaST-Scheduler - these tools work in tandem to optimize the performance of the GPU sharing system. The FaST-Profiler is responsible for analyzing and understanding the resource requirements and usage patterns of different DL inference functions. This profiling information is crucial for making informed decisions about resource allocation. Heuristic Scaling Algorithm and Maximum Rectangle Algorithm - these algorithms are integral to the FaST-Scheduler. The Heuristic Scaling Algorithm is designed to dynamically adjust resource allocation in response to changing demands and workloads. This ensures that resources are used efficiently and that the system can adapt to varying levels of demand. The Maximum Rectangle Algorithm, on the other hand, is focused on optimizing the scheduling of functions. It aims to maximize the utilization of available resources by arranging function executions in the most efficient manner possible.

\subsection{Techniques allowing for function-as-a-service model serving}
This section describes all techniques that could be used for the purpose of serving multiple machine learning inference workloads. 

Kernel-as-a-Service (KaaS) \cite{kaas} is an approach that integrates with the Ray distributed computing framework to enhance the execution of stateless and stateful functions across a cluster of machines. KaaS introduces "kTasks," which are executed by a specialized KaaS executor. This executor is responsible for managing kTask code and caching objects from the object store. It utilizes a kernel cache to link CUDA libraries and prepare kernels for invocation. 

Additionally, the concept of Exclusive Tasks (eTasks) is introduced. These tasks run on dedicated workers with exclusive GPU control. They can cache state between invocations but may be terminated to free up resources for new eTasks. eTasks offer a similar interface to popular FaaS platforms like AWS Lambda \cite{Lambda} or Google Cloud Functions \cite{CloudRun}.

Overall, KaaS effectively decouples GPUs from hosts, creating Directed Acyclic Graphs (DAGs) for GPU execution of tasks. Once a kTask is ready, it is executed in a pool of GPU workers as part of a DAG, and the results are subsequently returned. This approach represents a significant advancement in distributed computing, particularly in the efficient utilization of GPU resources.

Some works highlight that standard containerization frameworks like Docker \cite{Docker} are insufficient for the purpose of serving ml models for inference tasks. The initialization time of docker containers significantly exceeds the inference time of the model on GPU making container initialization the bottleneck of the pipeline. 

The Faaslet framework described in \cite{faaslets} presents a notable advancement over traditional containerization platforms like Docker, primarily in its approach to initialization, virtualization, and resource efficiency. One of the key features of the Faaslet framework is its use of Proto-Faaslets, which are essentially pre-initialized snapshots. This approach significantly mitigates the cold-start problem common in serverless computing. While traditional methods may take longer to initialize due to the need to set up a full environment, Proto-Faaslets can be restored quickly across hosts, enabling rapid horizontal scaling on clusters. This results in initialization times being reduced to just hundreds of microseconds.

In contrast to the standard POSIX environment used in containers, Faaslets employ a minimal virtualization layer. This layer is coupled with a low-level host interface that is specifically designed for high-performance serverless applications. The interface is dynamically linked with function code at runtime, making calls to the interface more efficient than performing similar tasks through an external API. This minimal virtualization approach significantly reduces the overheads required for isolation compared to traditional serverless platforms.

The work discussed in \cite{Clockwork} advocates for maximizing GPU utilization by limiting the choices available to lower system layers. This approach is based on the observation that performance variability often arises when lower layers in the system stack are given too many choices in executing tasks. The concept is applied across various layers, including hardware, operating system, and application levels. Hardware Level Choices - when a GPU is given multiple CUDA kernels to execute in parallel, it must decide how to allocate resources like execution units and memory bandwidth. These decisions are influenced by the GPU's internal state and proprietary policies, leading to unpredictability. Operating System Level Choices - creating multiple threads for execution on the same core gives the OS the choice of scheduling these threads based on its internal policies and state, leading to variability in execution. Application Level Choices - when distributed application worker processes manage their own caches or implement their own thread pools and queuing policies, they decide what to cache, for how long, and which requests to execute first. This leads to unpredictable hit rates, latency variability, and queuing times.

To address these issues, the authors proposed an architecture, named "Clockwork" which centralizes control and aims for predictable performance in workers. It implements a centralized controller - users submit inference requests to a central controller. This controller queues the requests and has a global view of the system state, including all workers. Each worker holds a set of DNN models in RAM and maintains exclusive control over one or more GPUs. The centralized scheduler decides when to execute each request, including when to load models into GPU memory and when to execute requests on the GPU. The scheduler can make accurate decisions regarding caching, scheduling, and load balancing because execution on workers is highly predictable. Workers execute schedules exactly as directed by the controller, which transmits continual scheduling information.

In \cite{gpu-enabled-faas} authors implement a scheduler, cache manager, and GPU manager which enable and optimize GPU functions upon an existing FaaS framework such as OpenFaaS. The Scheduler employs a Locality-Aware and Load-Balancing (LALB) approach, effectively managing GPU workloads and improving utilization. It treats frequently used models as cache items, reducing load times and enhancing performance. The Cache Manager oversees the management of models on the GPU, streamlining access and execution of inference tasks, especially for models that are regularly used. Lastly, the GPU Manager allocates and manages GPU resources, ensuring optimal utilization and balancing the demands of various inference tasks.

The work in \cite{trims} describes TrIMS, a memory sharing technique, facilitating the sharing of constant data across different processes or containers, ensuring user isolation. It incorporates a persistent model store that spans across GPU, CPU, local, and cloud storage. This system includes an effective resource management layer for maintaining isolation and a set of abstracts, APIs, and container technologies. These components are designed for seamless integration with FaaS platforms, DL frameworks, and user applications, offering a straightforward and transparent user experience.

The work described in \cite{storage} introduces DSCS-Serverless, a novel system architecture that leverages a computational storage drive (CSD) and a near-storage domain specific accelerator (DSA) to execute serverless functions directly on the storage node. This approach significantly reduces network and I/O data transfer overheads, enhancing resource utilization while complementing traditional disaggregated storage nodes. While originally not targeted for inference workloads, the idea could be expanded for multi model serving systems.

\section{Discussion and Reflection}
The strategy for efficiently managing GPU resources in machine learning model inference services, through dynamic model swapping and memory caching, tackles the problem of fluctuating user demands. This approach is a significant step away from the traditional idea of allocating separate instances for each model, which often leads to underutilization, especially for infrequently used models.

One of the primary challenges in this domain has been the increasing size of machine learning models, some occupying over 10GB of GPU memory. This increase has turned the loading times of these models into a critical bottleneck, affecting the service's ability to meet established SLOs and maintain user satisfaction. To address this, innovations like caching model parameters in RAM have been proposed and implemented. This method significantly reduces loading times by bypassing the slower process of retrieving data from disks, demonstrating how technical advancements can be harnessed to enhance system performance.

In the realm of dynamic swapping and caching strategies, systems like ModelMesh and the concept of using GPU memory as a cache for frequently used models are groundbreaking. These strategies optimize resource utilization by keeping regularly used models readily available in the GPU memory, thus reducing the time it takes to load these models when they are needed. This approach is particularly beneficial in environments where model usage patterns vary significantly among users.

The use of advanced hardware technologies like PCIe and NVLink for model swapping is another area where significant improvements have been made. These technologies offer much higher bandwidths compared to traditional methods, enabling quicker data transfers and more efficient model swapping processes. Such hardware advancements provide a path to overcoming the challenges posed by large model sizes and the need for rapid access to different models.

Concurrent operation strategies, such as overlapping model weight loading with GPU kernel executions, have shown to be effective in maximizing GPU usage. By loading the next set of parameters while the current layer is being processed on the GPU, these strategies significantly streamline the inference process, reducing latency times and enhancing overall system efficiency.

Resource sharing and memory management techniques, including the concept of sharing model tensors across replicas and employing fine-grained synchronization mechanisms like “cuSync”, are crucial in enhancing the efficiency of parallel computing tasks on GPUs. These approaches not only optimize memory usage but also contribute to a more efficient and cost-effective use of GPU resources.

The adoption of Just-in-Time (JIT) compilers, which dynamically allocate resources based on runtime information, marks a shift towards more adaptive and responsive resource management strategies. These compilers, by adjusting to the workload running on the GPU in real-time, ensure optimal performance and resource utilization.

In the context of optimizing GPU allocation for hosting multiple models, the use of technologies like Nvidia’s Multi-Instance GPU (MIG) to partition GPUs into smaller virtual instances has been a major improvement in the field. This technology allows for the simultaneous hosting and execution of multiple models on a single GPU, enhancing overall GPU utilization and adapting to varying traffic patterns and workload requirements.

Batching strategies, where requests are accumulated before execution, have also proven effective, especially in scenarios where immediate responses are not critical. Systems that employ batching techniques, like MLProxy and DNNScaler, illustrate how the balance between batching and response time requirements can be optimized for better system efficiency.

Lastly, exploring other resource optimization techniques like space-time scheduling, kernel merging, and NETFUSE indicates a continuous search for innovative ways to optimize the performance of concurrent model executions. These techniques aim to maximize GPU resource utilization, reduce latency, and enhance the overall efficiency of machine learning model inference services.

\section{Conclusion and Future Work}
The exploration and implementation of various strategies for managing GPU resources in the field of machine learning model inference have demonstrated a comprehensive understanding of the complexities and opportunities present in this rapidly evolving area. The integration of hardware advancements and innovative software solutions has paved the way for more efficient, responsive, and cost-effective services. These strategies, as discussed, have significantly addressed the challenge of fluctuating user demands and the efficient allocation of GPU resources.

A recurring theme in the majority of the cited works is the identification of a critical bottleneck: the transfer of models to the GPU for inference tasks. This transfer process often results in periods of GPU idle time, as the system waits for model weights to be loaded into the GPU memory. Addressing this inefficiency has been a focal point of recent research and development efforts.

Looking to the future, one promising direction is the incorporation of deep recurrent neural networks (RNNs) to predict the time of incoming user requests. The ability to forecast when a specific user is likely to require request processing could revolutionize the way model transfers to GPUs are managed. By predicting user request patterns, it becomes feasible to initiate the transfer of models to the GPU ahead of time. This preemptive approach could drastically minimize the waiting time for model transfers, ensuring that the GPU is ready to perform inference tasks as soon as a new request is received.

Implementing deep RNNs for this purpose could lead to a significant reduction in response times and an increase in overall system efficiency. This strategy would not only optimize GPU utilization but also enhance user satisfaction by providing faster and more reliable service.

The strategic management of GPU resources in machine learning model inference, through dynamic model swapping and memory caching, marks a significant evolution from traditional methods. Addressing challenges like large model sizes and lengthy loading times, enhanced system performance, and user satisfaction. Innovations such as using advanced hardware technologies, concurrent operation strategies, and resource sharing techniques have optimized GPU utilization. As we continue to explore new optimization techniques, the potential for further advancements in this dynamic domain remains vast, promising even more efficient and effective machine learning model inference services in the future.

{\small
\bibliographystyle{ieee}
\bibliography{bibliography}
}

\end{document}